\UseRawInputEncoding
\documentclass[twocolumn,secnumarabic,amssymb, nobibnotes, aps, prx, 10pt, floatfix]{revtex4-2}
\usepackage{amsmath, amssymb}
\usepackage{color, hyperref}
\usepackage[all]{hypcap} 
\usepackage{graphicx}
\usepackage{xcolor}
\usepackage{placeins}
\usepackage[normalem]{ulem}
\usepackage{bm}
\usepackage{tabularx}
\usepackage{multirow}

\newcommand{\Ket}[1]{ \mbox{$|#1\rangle$} }
\newcommand{\ME}[3]{ \mbox{$\langle #1\,|\,#2\,|\,#3\rangle $} }

\usepackage{graphicx}
\usepackage{dcolumn}
\usepackage{bm}
\usepackage{amsmath}
\usepackage{color}
\usepackage{hyperref}
\hypersetup{
    colorlinks=true,
    linkcolor=blue,
    citecolor=blue,
    filecolor=blue,      
    urlcolor=blue,
}
\usepackage{cleveref}
\usepackage[normalem]{ulem}

\newcommand{\STAB}[1]{\begin{tabular}{@{}c@{}}#1\end{tabular}}

\begin{document}

\title{Role of intermediate resonances in attosecond photo\-electron inter\-ferometry in neon}

\author{M.~Moioli$^{1}$, M.~M.~Popova, K.~R.~Hamilton$^{2,3}$, D. Ertel$^{1}$, D. Busto$^{1,4}$, I.~Makos$^{1}$, 
M.~D.~Kiselev, S.~N.~Yudin, H.~Ahmadi$^{1}$, C.~D.~Schr\"{o}ter$^{5}$, T.~Pfeifer$^{5}$, R.~Moshammer$^{5}$, E.~V.~Gryzlova, 
A.~N.~Grum-Grzhimailo, K.~Bartschat$^{2}$, and G.~Sansone$^{1,6,*}$}

\affiliation{$^{1}$ Physics Department University of Freiburg, 79104 Freiburg, Germany}
\affiliation{$^{2}$ Department of Physics and Astronomy, Drake University, Des Moines, Iowa 50311, USA}
\affiliation{$^{3}$ Department of Physics, University of Colorado Denver, Denver, Colorado 80204, USA}
\affiliation{$^{4}$ Department of Physics,
Lund University, SE-221 00 Lund, Sweden}
\affiliation{$^{5}$ Max-Planck-Institut f\"{u}r Kernphysik, 69117 Heidelberg, Germany.}
\affiliation{$^{6}$ Freiburg Institute for Advanced Studies (FRIAS), University of Freiburg, 79104 Freiburg, Germany.}
\email{corresponding author: giuseppe.sansone@physik.uni-freiburg.de}

\date{\today}
\begin{abstract}
Atto\-second photo\-electron inter\-ferometry based on the combination of an attosecond pulse train and a synchronized infrared field is a fundamental technique for the temporal characterization of attosecond waveforms and for the investigation of electron dynamics in the photo\-ionization process.
In this approach, the comb of extreme ultraviolet harmonics typically lies above the ionization threshold of the target under investigation, thus releasing a photo\-electron by single-photon absorption. The interaction of the outgoing photo\-electron with the infrared pulse results in the absorption or emission of infrared photons, thereby creating additional peaks in the photo\-electron spectrum, referred to as sidebands. While, in the absence of resonances in the first ionization step, the phases imparted on the photo\-ionization process evolve smoothly with the photon energy, the presence of intermediate resonances imprints a large additional phase on the outgoing photo\-electron wave packet.
In this work, using a comb of harmonics below and above the ionization threshold of neon, we investigate the effect of intermediate bound excited states on atto\-second photo\-electron interferometry. We show that the phase of the oscillations of the sidebands and their angular distributions are strongly affected by such resonances. By slightly tuning the photon energies of the extreme ultraviolet harmonics, we show how the contributions of selected resonances can be enhanced or suppressed. 
\end{abstract}
\maketitle
 
\section{Introduction}
The reconstruction of atto\-second beating by inter\-ference of two-photon transitions (\hbox{RABBIT}) technique was demonstrated in 2001~\cite{Science-Paul-2001b}, with the goal to reconstruct the temporal profile of an atto\-second pulse train. The oscillations in the photo\-electron spectra for the two-color photo\-ionization process as a function of the relative delay $\tau$ between the extreme ultra\-violet (XUV) and infra\-red (IR) pulses contain information about the phase difference between consecutive harmonics, thereby giving access to the atto\-second chirp~\cite{Science-Mairesse-2003,JMO-Varju-2005}. This information can be extracted from the experimental data, provided that the target-specific phases imprinted on the photo\-electron wave packets are known and can be subtracted from the measured phase differences. In the case of XUV photon energies well above the ionization threshold, and in the absence of resonances in the continuum, these phases are expected to evolve smoothly with the photon energy. In atoms, they can be further decomposed into a photo\-ionization phase, associated with the photo\-ionization process due to the absorption of a single XUV photon, and a continuum-continuum phase due to the interaction of the outgoing photo\-electron wave packet with the IR field \hbox{\cite{Dahlstr_m_2012,Dahlstr_m_2014,RevModPhys.87.765}}.
More recently, the validity of this ``decomposition approximation'' was generalized and tested also for multiple continuum-continuum transitions~\cite{PhysRevA.103.022834,PhysRevA.107.022801,PhysRevA.109.023110}.

The presence of intermediate resonances, however, either in the continuum or in the bound-state spectral region, is expected to play a crucial role in atto\-second photo\-electron inter\-ferometry, as these levels can modify the amplitude and phase of one or even both of the two pathways contributing to the inter\-ference signal~\cite{PRA-Kheifets-2021}.
These effects have been investigated in helium, addressing the role of intermediate bound states reached by single XUV-photon ionization~\cite{PRL-Swoboda-2010, SCIADV-Autuori-2022,PRA-Drescher-2022,FP-Neoricic-2022} and of doubly-excited states in the continuum~\cite{SCIENCE-Gruson-2016}. 
In helium, the large energetic distance between consecutive excited levels accessible by single-photon excitation from the ground state, compared to the width of a single harmonic, makes it possible to address individual resonances for each configuration of the harmonics.

In heavier noble-gas atoms on the other hand, the manifold of excited states accessible by dipole transitions from the ground state becomes energetically closer~\cite{villeneuveCoherentImagingAttosecond2017}. A single harmonic may thus excite more than one level, thereby generating a rich interference structure in the photo\-electron signal in the continuum. Recently, the effect of an intermediate resonance in the interference between two pathways determined by XUV colors was also investigated~\cite{NATPHYS-Carpeggiani-2019, PRR-Gryzlova-2022}. In molecules, intermediate shape resonances can also affect the phase of the photo\-emission measured in the laboratory~\cite{SCIADV-Nandi-2020} and in the recoil frame~\cite{SCIADV-Heck-2021, NATCOMM-Ahmadi-2022} of the target system.

\begin{figure}[t!]
\centering 
\includegraphics[width=0.47\textwidth]{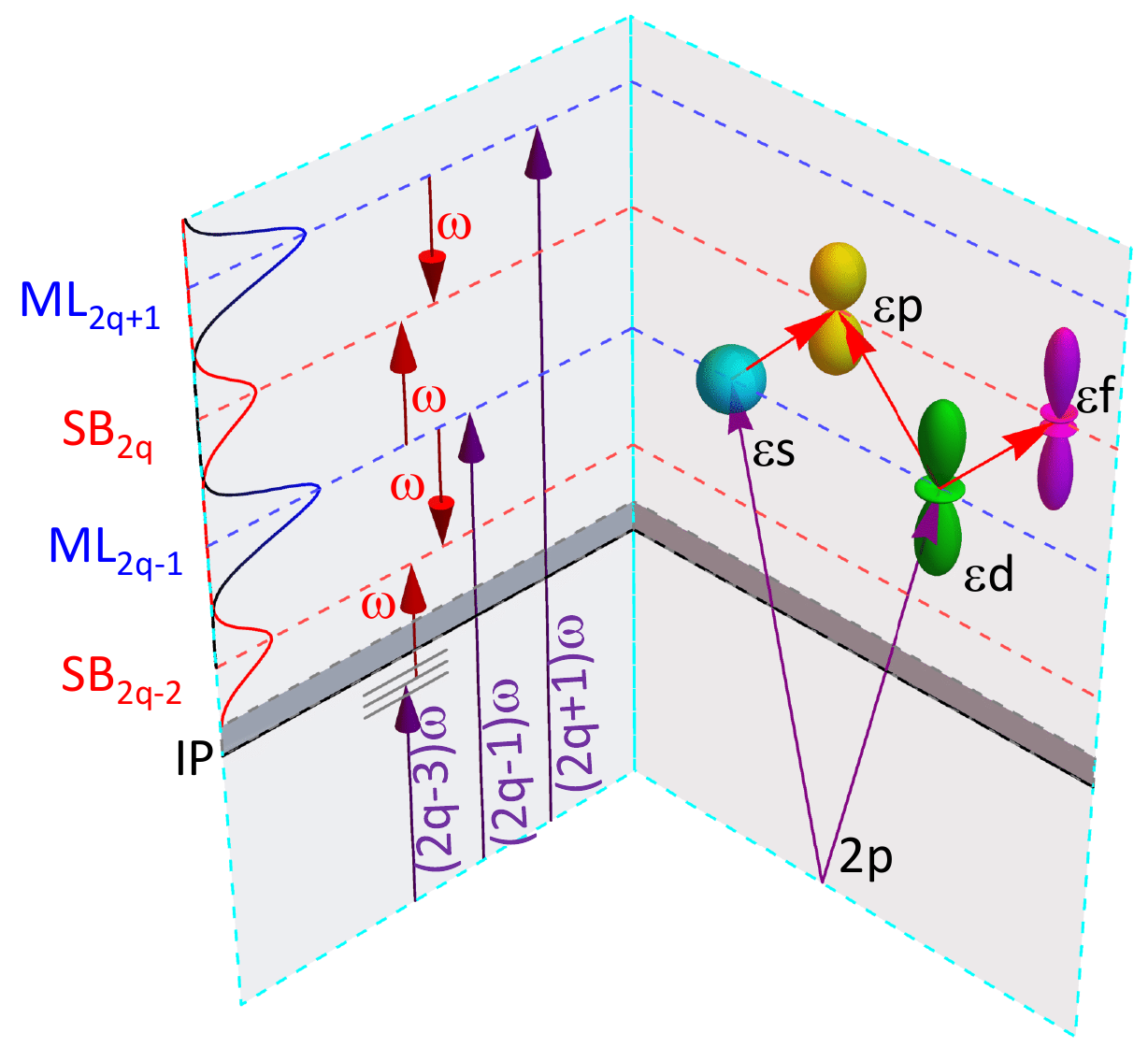}
\caption{Scheme of the energy levels and angular character of the photo\-electron wave packets emitted by a comb of odd XUV harmonics around the ionization threshold of neon. The UT-\hbox{RABBIT}  mechanism is formed by the harmonics $2q-3$ and $2q-1$, while the pairs $2q-1$ and $2q+1$ form the AT-\hbox{RABBIT} SB$_{2q}$ line. The angular distributions on the right-hand-side describe schematically the photo\-electron with kinetic energy~$\varepsilon$ in the continuum. These electrons are ejected after absorption of a single XUV photon, resulting in $\varepsilon s$ and $\varepsilon d$ states. Their partial-wave character is then further modified by the absorption (or emission) of an additional IR photon, 
resulting in $\varepsilon p$ and $\varepsilon f$ states.} 
\label{Fig1}
\end{figure}

The scheme of the present study is presented in Fig.~\ref{Fig1}. A comb of XUV harmonics generated by high-order harmonic generation (HHG) in gases includes harmonics below (harmonic $2q-3$) and above (harmonics $2q-1$ and $2q+1$) the ionization threshold of neon. While the harmonic below the ionization threshold can populate a resonant bound excited state, the absorption of a single XUV photon above the ionization threshold leads to the direct emission of a photo\-electron (indicated as main lines ML$_{2q-1}$ and ML$_{2q+1}$). As a consequence, the first sideband above the ionization threshold (SB$_{2q-2}$) is formed by the interference of one path influenced by the presence of a manifold of bound states below the ionization threshold and another one resulting from non-resonant ionization above the ionization threshold. 
We label the process leading to the formation of the sideband $S_{2q-2}$ as ``under-threshold" (UT)-\hbox{RABBIT}~\cite{PRA-Kheifets-2021,Kheifets_2023}, because of the involvement of the bound state below the ionization threshold. In contrast, the ``above-threshold'' (AT)-\hbox{RABBIT}  mechanism occurs when both XUV absorption processes already lead to the continuum. 

The absorption of a single XUV photon resulting in an electron ejected from the $2p$ sub\-shell of neon leads to the population of continuum states with $s$ and $d$ character, 
i.e., orbital angular momenta $\ell=0$ and $\ell=2$, respectively~\cite{josephAngleresolvedStudiesXUV2020}. The subsequent absorption or emission of an IR photon will result in population of final states with \hbox{$p$ ($\ell=1)$} and \hbox{$f$ ($\ell=3$)} character. The right-hand-side of Fig.~\ref{Fig1} shows schematically the character of the different states populated by the absorption of an XUV and IR photon above the ionization threshold of neon. The same conclusions are valid when the first XUV photon populates a bound state below the ionization threshold.


\section{Experimental Setup}\label{Setup}
The experimental campaigns were performed using two experimental atto\-second sources described in detail in refs.~\cite{JPP-Ahmadi-2020,RSI-Ertel-2023}. 
For the first one, the laser driving the HHG was a titanium-doped sapphire system operating at 10~kHz, with a central wavelength of $\lambda=800$~nm delivering pulses with an energy of 1~mJ and a duration (FWHM of intensity profile) of 30~fs. The second one produced IR pulses starting from a Yb:KGW regenerative amplifier with a central wavelength of 1025~nm, 50~kHz repetition rate, and 400~$\mu$J pulse energy. The original pulse duration of 277~fs was reduced to 17.8~fs by exploiting nonlinear broadening in a gas-filled hollow-core fiber (HCF) and subsequent temporal compression using chirped mirrors.~\cite{RSI-Ertel-2023}. 
In the HCF, the type and the pressure of the gas lead to different blue shifts of the central wavelength, thereby offering the possibility to slightly tune the central wavelength in the range between 1001~nm and 1019~nm with an uncertainty of about $\pm 5$~nm.
In both beamlines, a delay between the IR pulses and the XUV attosecond pulse train was introduced and controlled with an ultra-stable collinear setup based on drilled glass plates~\cite{JPP-Ahmadi-2020, RSI-Ertel-2023}.

The XUV and IR pulses were focused by a toroidal mirror in a 3D-momentum-imaging spectrometer (reaction microscope)~\cite{Ullrich_2003}, where they interacted with a supersonic neon gas jet. The products of the ionization were then guided to the photo\-electron and photo\-ion detectors by suitable electric and magnetic fields.
Here, the peak intensities of the IR pulses at the interaction region of the reaction microscope were estimated as $I_{{\rm IR}}\approx 1\times 10^{12}\,\rm W/cm^2$ and $I_{\rm IR}\approx 7\times 10^{11}\,\rm W/cm^2$ for the first and second experimental setup, respectively. At the same position, the highest peak intensity of a single attosecond pulse in the train for the two setups was $I_{\rm XUV} \approx 2 \times 10^{10}\,\rm W/cm^2$ and $I_{\rm XUV} \approx 5 \times 10^{9}\,\rm W/cm^2$, respectively.

The spectral distribution of the XUV harmonics was characterized using an XUV spectrometer placed after the coincidence spectrometer. The duration of each harmonic was estimated as $\approx 9~$fs for 800~nm and $\approx 6~$fs for 1006~nm by considering the width of the corresponding spectral peaks, the XUV spectrometer resolution, and assuming a gaussian envelope. 

The experimental parameters for the IR and XUV fields used in our experiments are summarized in Table~\ref{table:sppl}. The strength of the electric field of each harmonic is given relative to the 19$^{\rm th}$ harmonic for 800~nm and the 25$^{\rm th}$ harmonic for 1006~nm, respectively. The  lowest  (11$^{\rm th}$ for 800~nm and 13$^{\rm th}$-15$^{\rm th}$ for 1006~nm) and the highest (23$^{\rm rd}$ for 800~nm and 31$^{\rm st}$ for 1006~nm) harmonics were outside the spectral range of the XUV spectrometer. Their intensities are assumed to be equal to the closest harmonic measured in the experiment. The phases of the lowest harmonics (including those contributing to the \hbox{UT-RABBIT}) and highest harmonics were determined assuming a constant value of the group delay dispersion (GDD) determined from the oscillations of the closest \hbox{AT-RABBIT} sidebands.

\begin{table}[t!]\label{table:sppl}
\centering
\caption{Parameters of the polychromatic XUV field. The error values of the amplitudes ${E}_{2q+1}$ are calculated as the standard deviation of the average values around each harmonic maximum.	For the phases ${\phi}_{2q+1}$, the error values are evaluated by weighting the standard deviation of the phases around the harmonic maxima with the corresponding amplitude values around the peak.}

\medskip

\begin{tabular}{c|c|c|c}
\hline
& Harmonic order $2q+1$ & $E_{2q+1}$  & $\phi_{2q+1}$ (rad)\\
\hline \hline
 \multirow{7}{*}{\STAB{\rotatebox[origin=c]{90}{$\lambda=800$ nm}}}
&11 & $ 0.37 \pm 0.02$  &  $\phantom{-}0.00 \pm0.02$  \\
&13 & $ 0.37 \pm 0.02$  &   $\phantom{-}0.79 \pm 0.02$ \\
&15 & $0.68 \pm 0.01$  &    $-1.93 \pm 0.02$ \\	
&17 & $0.93 \pm 0.01$  &    $-3.46 \pm0.02 $ \\	
&19 & $1.00 \pm 0.01$  &    $\phantom{-}0.85 \pm 0.07$ \\	
&21 & $0.82 \pm 0.01$  &     $-1.51 \pm 0.39$ \\	
&23 & $0.82 \pm 0.01$  &       $-4.13 \pm 0.39$ \\	
\hline
\hline
 \multirow{10}{*}{\STAB{\rotatebox[origin=c]{90}{$\lambda=1006$ nm}}}
&13 & $0.59 \pm 0.01$ &   $\phantom{-}0.00 \pm0.01$  \\
&15 & $0.59 \pm 0.01$ &   $\phantom{-}0.04 \pm 0.01$ \\	
&17 & $0.59 \pm 0.01$ &   $-0.23 \pm 0.01$ \\	
&19 & $0.86 \pm 0.02$ &   $-0.78 \pm 0.05$ \\	
&21 & $0.97 \pm 0.01$ &   $-1.63 \pm 0.03$ \\	
&23 & $0.92 \pm 0.02$ &   $-2.72 \pm 0.08$ \\	
&25 & $1.00 \pm 0.01$ &   $\phantom{-}2.23 \pm 0.03$ \\	
&27 & $0.99 \pm 0.01$ &   $\phantom{-}0.48 \pm 0.07$ \\	
&29 & $0.85 \pm 0.01$ &  $-1.56 \pm 0.1$  \\	
&31 & $0.85 \pm 0.01$ &    $\phantom{-}2.39 \pm 0.1$ \\
\hline
\end{tabular}
\end{table}

We consider the \hbox{RABBIT} scheme for the angle-integrated electron emission probability as well as photo\-electron angular distributions (PADs):
\begin{eqnarray} \label{eq:PAD}
 S(\theta,\tau) 
  &=& S_0(\theta)+S_2(\theta)\cos{(2\,\omega_{\rm IR}\tau-\phi(\theta))}.
\end{eqnarray}
Here, $\theta$, $\omega_{\mathrm{IR}}$, and $\tau$ indicate the ejected-electron angle defined with respect to the polarization axis of the fields, the frequency of the IR field, and the relative delay between the atto\-second pulse train and the IR field, respectively. The phase $\phi(\theta)$ is the phase introduced by the photoionization process, which, in general, depends on the emission angle of the photoelectron. By averaging the delays $\tau$ over an integer number of periods~$T$ of the IR, the PAD can be described using the angular anisotropy parameters $\beta_n$ and the Legendre polynomials $P_n$ according to the formula:
\begin{eqnarray} \label{eq:PADaveraged}
 S_0(\theta) &=&\frac{S_0}{4\pi}
  \left[ 1+\beta_2P_2(\cos\theta)+\beta_4 P_4(\cos\theta)\right].
\end{eqnarray}
Integrating Eq.~(\ref{eq:PAD}) over the angles yields 
\begin{eqnarray}\label{eq:Klaus} 
 S_0(\tau) &=&
   S_0+S_2\cos{(2\,\omega_{\rm IR}\tau-\phi)}.
\end{eqnarray}
In addition to the $(\theta,\tau)$ dependence written explicitly, all observable values are functions of the photo\-electron energy $\varepsilon$, which is omitted for brevity of notation.

Both the \hbox{RABBIT} phase and the $\beta_n$ parameters can be extracted by fitting experimental data and theoretical predictions to the above forms.

\section{Theoretical models} \label {Theory}
In this section, we summarize the theoretical models used to predict the observables of interest. 
They are:
\begin{itemize}
    \item the non\-relativistic R-matrix with time dependence (RMT) approach,
    \item a relativistic time-dependent perturbation \hbox{theory} (PT), and
    \item an approach based on solving rate equations (RE).
\end{itemize} 

We consider an electromagnetic field consisting of odd harmonics $\omega_{2q+1}$ of the fundamental radiation with frequency $\omega_{\rm{IR}}$ with phases $\phi_{2q+1}$ and envelopes $\bm{E}_{2q+1}$ characterized by a duration $\sigma_{2q+1}$, 
together with an IR field with phase $\phi_{\rm{IR}}$ and envelope $\bm{E}_{\rm{IR}}$ characterized by a duration $\sigma_{\rm{IR}}$:
\begin{eqnarray}
 \label{eq:field}
	\bm{E}(t)&=&\sum_{q}\bm{E}_{2q+1}[t,\sigma_{2q+1}] \cos(\omega_{2q+1} t+\phi_{2q+1}) +\nonumber\\
 &+&\bm{E}_{\rm{IR}}[(t-\tau),\sigma_{\rm{IR}}]\cos(\omega_{\rm{IR}} (t-\tau)+\phi_{\rm{IR}}).
\end{eqnarray}
The temporal shape of the envelopes is assumed to be the same for all  XUV harmonics $\bm{E}_{2q+1}$. Their amplitudes are scaled according to the relative strengths shown in Table~\ref{table:sppl}.
In the RMT calculations, we read in the estimated experimental pulse, sampled on a numerical time grid.
In PT and RE, we fitted the pulse to the numerical form and then used $\sin^2$ envelopes to approximate the gaussians, since 
this is advantageous for evaluating some of the necessary integrals analytically. 

As in the experimental setup, all harmonics are linearly polarized along the quantization axis $\bm{z}$ of the atomic system. 

\subsection{R-matrix with Time Dependence (RMT)} \label {RMT}
RMT is an {\it ab-initio} technique that solves the time-dependent Schr\"odinger equation by employing  the \hbox{$R$-matrix} paradigm, which divides the configuration space into two separate regions over the radial coordinate of the ejected electron~\cite{RMT}. A general computer code was published~\cite{RMT_CPC}, with the most recent version available on GitLab~\cite{RMTgit}.

In the inner region, close to the nucleus, the time-dependent $N$-electron wave function is represented by a time-independent  $R$-matrix (close-coupling) basis with time-dependent coefficients. In the outer region, the wave function is expressed in terms of residual-ion states coupled with the radial wave function of the ejected electron on a finite-difference grid. The solutions in the two separate regions are then matched directly at their mutual boundary. The wave function is propagated in the length gauge of the electric dipole operator. For the typical atomic structure descriptions used in RMT, this converges more quickly than the velocity gauge~\cite{tdrm_dipole_gauge}.  On the other hand, propagation in the length gauge usually requires more angular momenta to obtain partial-wave-converged results~\cite{Cormier_1996,PhysRevA.81.043408}. 

In the non\-relativistic RMT calculations carried out for the present project, the neon target was described within an $R$-matrix inner region of radius 20 (we use atomic units in Sec.~\ref{Theory}) and an outer region of 15,000. The finite-difference grid spacing in the outer region was 0.08 and the time step for the wave-function propagation was 0.01.  

The target was described using the one-electron orbitals generated by Burke and Taylor~\cite{window_resonances_argon} in their 2-state model, which couples the ionic ground state $(2s^2 2p^5)^2P$ and the first excited state~$(2s 2p^6)^2S$.
Multi\-configuration expansions, including pseudo-orbitals to account for the
term dependence of the physical $2s$ and $2p$ orbitals, were employed for both ionic states as well as the initial $(2s^2 2p^6)^1S$ state.

We then included all available $2s^22p^5\epsilon \ell$ and $2s2p^6\epsilon \ell$ 
channels up to a maximum total orbital angular momentum of $L_{\mathrm{max}}=\mathrm{30}$.
The continuum functions were constructed from a set of 50 \hbox{$B$-splines} of order 13 for each angular momentum of the outgoing electron. 

In order to extract the transition amplitudes, the final-state radial wavefunction of the ejected electron was projected onto Coulomb functions.  This is an improvement over the original formulation of RMT, where the projection was performed onto plane waves, i.e., Bessel functions. 
 While the change had some effect, it did not alter the qualitative nature of the present predictions.  Consequently, it seems unlikely that an even more appropriate projection onto distorted Coulomb waves is necessary in light of the other approximations already made in the model.
 
\subsection{Rate-Equation Approach (RE) and Time-Dependent Perturbation Theory (PT)} 
Two methods based on the same spectroscopic model were developed: direct numerical solution of a system of rate equations (RE) and non-stationary perturbation theory (PT). 
Employing two approaches based on the same spectroscopy allows us to clearly distinguish various strong-field effects, such as the Stark shift. 

In the RE approach, the wave function of a state is represented as an expansion in a basis of unperturbed atomic wave functions $\psi_n(\bm{r})$:
\begin{equation}\label{eq:wf}
    \Ket{\Psi(\bm{r},t)}=\sum_n a_{n}(t)\psi_n(\bm{r})\,,
\end{equation}
where the index $n$ denotes all quantum numbers needed to characterize the state: energy, orbital and total angular momentum, etc. The ``ion+photo\-electron" states are discretized with an energy step \hbox{$\delta \varepsilon=0.002$}. Substituting Eq.~(\ref{eq:wf}) into the Schr\"odinger equation in the velocity gauge, one obtains a system of coupled differential equations for the coefficients:
\begin{eqnarray}\label{eq:SRE}
\frac{d a_n(t)}{dt} &\!=\!& -\frac{\textrm{i}}{c}\hskip1truemm \int\hskip-5truemm \sum_{i} e^{\textrm{i}(\varepsilon_n-\varepsilon_i)t} \ME{\psi_n}{\bm{A}(t)\cdot\bm{p}}{\psi_i} \, a_i(t).  
\end{eqnarray}

Equation~(\ref{eq:SRE}) may be directly solved numerically \textcolor{black}{and, in the following, we indicate this approach as Rate Equations (RE) for the amplitudes method}. \textcolor{black}{Equation~(\ref{eq:SRE}) can also serve} as the starting point for non-stationary perturbation theory: to begin with, one sets \hbox{$a_n^{(0)}(t)=\delta_{n,1}$}, which means that initially the system is in the ground state. Then Eq.~(\ref{eq:SRE}) is integrated to find the first-order expansion coefficients $a_n^{(1)}(t)$.  Next, these coefficients are put back into Eq.~(\ref{eq:SRE}) to obtain the second- and even higher-order approximations. \textcolor{black}{In the following, we refer to this second approach as perturbation theory (PT)}.
\textcolor{black}{In PT, the solution of Eq.~(\ref{eq:SRE}) is expanded into a series as:
\begin{equation}\label{series}
a_n(t)= a_n^{(0)}+ a_n^{(1)}+ a_n^{(2)}….
\end{equation}
Even- and odd-rank $a_n^{(j)}$ have maxima at different photoelectron energies corresponding to SBs (even $j$) and MLs (odd $j$). While the sideband oscillations observed in this study are determined by the interference of second-order amplitudes, oscillations in the mainlines arise due to the interference of first- and third-order amplitudes. The probability for finding an electron in any particular state is defined by the square of the absolute amplitudes. Note that the square of $a_n^{(2)}$ and the interference product of $a_n^{(1)}$ and $a_n^{(3)}$ are of the same order.} 

The dipole transition discrete-discrete and discrete-continuum matrix elements were obtained in the $jK$-coupling scheme by employing Zatsarinny's \hbox{$B$-spline} \hbox{$R$-matrix} (BSR) code~\cite{zatsarinny2006}, which can be run in electron-ion collision mode with appropriately modified boundary conditions. In this coupling scheme, the quantum numbers characterizing a state are the total electronic angular momentum $J_f$ of the ionic core $(2p^5)^2P_{J_f}$, the orbital angular momentum~$l$ of the electron in the discrete $nl$ or continuum $\varepsilon l$ part of the spectrum, the intermediate quantum number $K$ being the result of the vector addition \hbox{$\bm{K}=\bm{J_f}+\bm{l}$}, and, finally, the total electronic angular momentum of the system \hbox{$\bm{J}=\bm{K}+\bm{1/2}$}, where $1/2$ is the electron spin. The choice of the $jK$-scheme made it possible to carefully take into account transitions from discrete to continuum states 
\textcolor{black}{with final ionic states that have fine-structure splittings of up to 0.1 eV for different values of $J_f$.}

The same spectroscopic model as in~\cite{popova2021} was used: all possible terms of configurations $1s^2 2s^2 2p^5ns$ ($n=3-6$), $1s^2 2s^2 2p^5nd$ ($n=3-5$), $1s^2 2s^2 2p^4 3s3p$, $1s^2 2s^2 2p^4 3p4s$, and $1s^2 2s^1 2p^5 3s3d$ were included, and the experimental ionization energies~\cite{NIST_ASD} were used.

The calculation of the continuum-continuum transition matrix elements was carried out under the assumption that any correlations between the photo\-electron and electrons of the residual ion can be neglected. After the self-consistent wave function of the $(2p^5)^2P$ ion was found with the MCHF software package~\cite{fischer1997}, the wave functions of the continuum electron in the frozen field of this ion 
were produced. Then, using the method described in~\cite{Mercouris1996}, the continuum-continuum radial matrix elements $R_{\varepsilon l, \varepsilon' l'}$ were calculated.
The relationship between $R_{\varepsilon l, \varepsilon' l'}$ and the reduced matrix elements in the $jK$-coupling scheme can be obtained using standard formulas from the quantum theory of angular momentum~\cite{polcor}.

 The results for the $d$- to $f$-wave transitions coincide with Gordon's formula ~\cite{Gordon1929}  to an accuracy of \hbox{$2-3$} percent.
 Being very asymmetric, these transitions proceed mainly by photon absorption and provide the essential contribution to the low-energy part of the spectrum.

\section{Results and Discussion} \label{Discussion}

\subsection{Delay-averaged data}

\begin{figure}[t!]
\centering 
\includegraphics[width=0.47\textwidth]{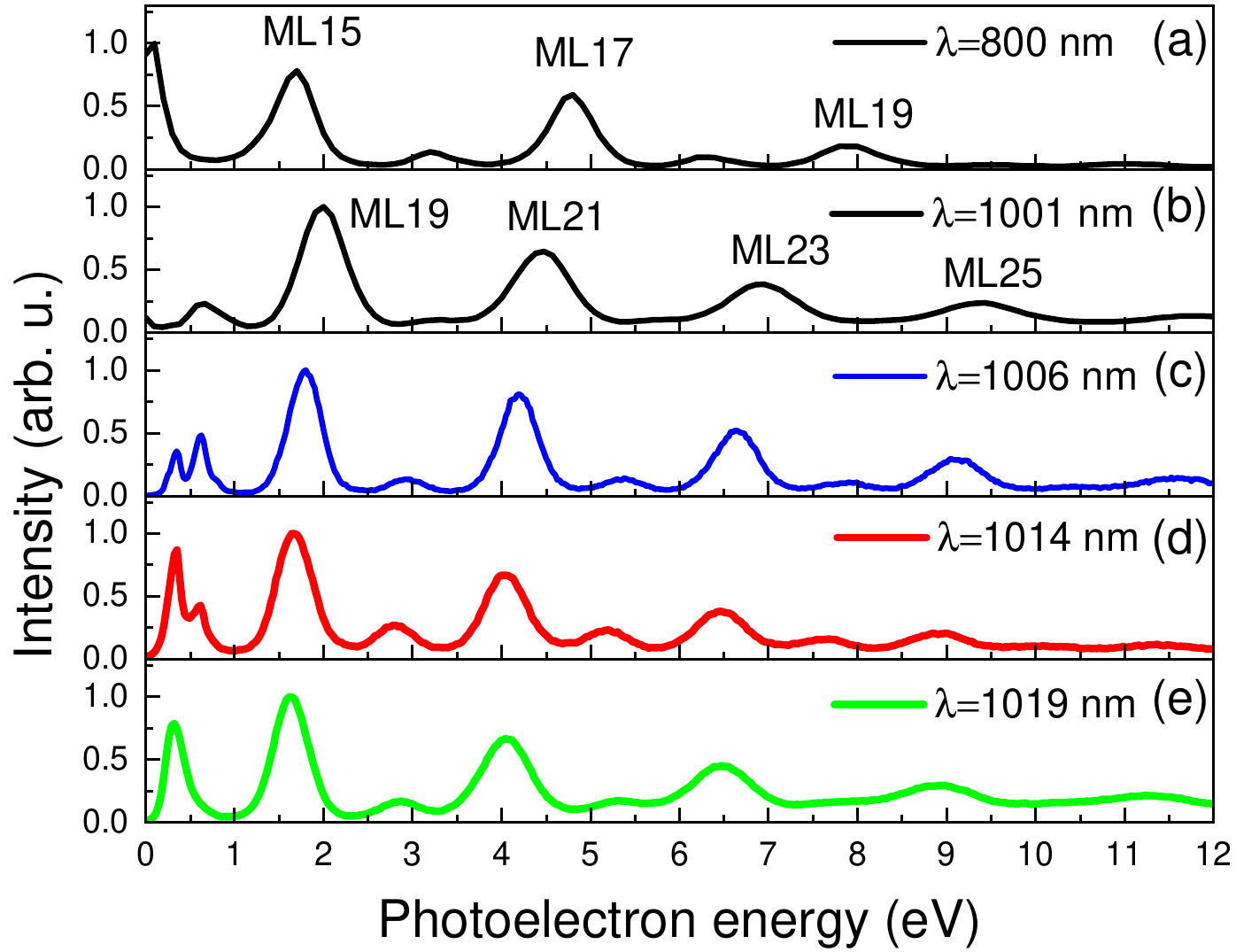}
\caption{
\hbox{RABBIT}  spectra integrated over the IR-XUV delay and acquired for different wavelenghts of the driving laser: $\lambda=800$~nm (a), $1001$~nm (b), $1006$~nm (c), $1014$~nm (d), and $1019$~nm (e). The UT-RABBIT sideband between 0 and 1~eV for the driving wavelengths in the range $1001-1019$~nm exhibits large variations, as seen in panels \hbox{(b)-(e)}. Note in particular the double-peak structure in panels~(c) and~(d), which is due to near-resonant excitation of at least two states (see text for details).
The uncertainty in the estimation of the driving wavelength is about $\pm 5$~nm.
}\label{Fig2}
\end{figure}

Photo\-electron spectra acquired with driving laser fields centered at \hbox{$\lambda=800$~nm} and  \hbox{$\lambda$} comprised between $1001$~nm and $1019$~nm are presented in Fig.~\ref{Fig2}.

The lowest energy sideband for the 800~nm wavelength is located around the ionization threshold extending up to about $0.2$~eV in the continuum. Its signal is stronger than that of the adjacent main line ML$_{15}$. This observation suggests that the sideband is mostly populated through a resonant pathway from below the ionization threshold with the absorption of a photon of the 13$^{\rm{th}}$~harmonic. Comparison with the tabulated data~\cite{NIST_ASD} strongly suggests that the resonant state is the $(2p^5 3d)^1P$,  which can be populated by single-photon absorption from the ground state.

\begin{figure*}   
\centering 
\includegraphics[width=0.99\textwidth]{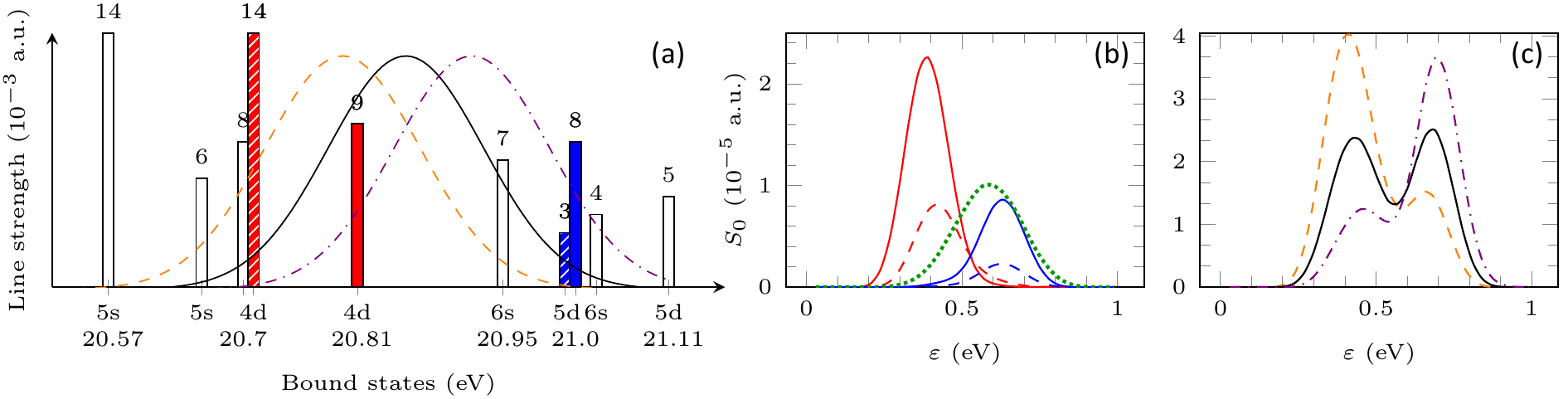}
\caption{
(a)~Energy position of the bound states reached by single-photon excitation starting from the ground state of neon with a driving wavelength of 1006~nm (black gaussian curve), 1008~nm (orange dashed curve), and 1004~nm (purple dash-dotted). The numbers above the columns correspond to the oscillator strength from the ground state.
(b)~Largest contributions to SB$_{18}$ in PT from discrete states for a driving wavelength of 1006~nm. The solid red and blue lines correspond to the red ($2p^5(^2P_{1/2})4d[3/2]$) and blue ($2p^5(^2P_{3/2})5d[3/2]$) full columns in panel~(a) while the red and blue dashed lines correspond to the red ($2p^5(^2P_{3/2})4d[3/2]$) and blue ($2p^5(^2P_{3/2})5d[1/2]$) hatched columns. The leading uncolored components of the $4d$ and $5d$ states are $2p^5(^2P_{3/2})4d[1/2]$ and $2p^5(^2P_{1/2})5d[3/2]$, respectively. The dotted green curve represents the contribution 
due to absorption of one photon of the $19^{\rm th}$ harmonic and subsequent emission of an IR photon from ML$_{19}$.  
(c)~Delay-averaged near-threshold spectra for 1006~nm (black), 1008~nm (orange dashed) and 1004~nm (purple dash-dotted) as obtained in the PT model. 
}
\label{Fig3}
\end{figure*}

The assignment of the intermediate resonance(s) is more complex in the second range of wavelengths ($1001$~nm$ \leq\lambda\leq1019$~nm), as the 17$^{\rm{th}}$~harmonic can reach a spectral region close to the ionization threshold, which is characterized by a high density of bound states. As mentioned previously, the central wavelength around 1010~nm can be slightly varied by changing the gas pressure in the HCF.  This offers the possibility of tuning the energy of the harmonic comb as shown by the four photo\-electron spectra presented in panels \hbox{(b)-(e)} of Fig.~\ref{Fig2}.  Interestingly, the characteristics of the first sideband are strongly affected by the precise central wavelength of the driving field. The side\-band is characterized by a single (panels~(b) and~(e)) or a double (panels~(c) and~(d)) peak, depending on the specific wavelength. This structure suggests that at least two intermediate bound states are involved in the population of the side\-band. 

For the detailed investigation of the effects close to the ionization threshold, photo\-electron spectra for a central wavelength of $\lambda\approx 1006$~nm were chosen, with special emphasis on optimizing the spectrometer resolution in the region of low kinetic energies. 

In order to identify the largest contributions from the excited resonances involved in the below-threshold ionization pathway, we carried out a series of calculations within the PT model. Specifically, we only ``switched on" one of the considered intermediate states and omitted all continuum-continuum transitions.
The calculated oscillator strengths for selected transitions between the ground and excited states are presented in Fig.~\ref{Fig3}(a).   The  largest contributions are presented in Fig.~\ref{Fig3}(b), together with the contribution due to absorption of one photon of the $19^{\rm th}$ harmonic and the subsequent emission of an IR photon from ML$_{19}$. 

The positions of the experimental maxima are in good agreement with the values expected due to ionization from the excited $(2p^5 4d)^1P$ state.  More precisely, although not resolvable in the experiment, from $2p^5(^2P_{3/2})4d[3/2]$ and $2p^5(^2P_{1/2})4d[3/2]$ as well as from $2p^5(^2P_{3/2})5d[3/2]$ and $2p^5(^2P_{3/2})5d[1/2]$ states to the final ionic states with total electronic angular momentum $J_f=1/2$ and $J_f=3/2$, which appear to contribute the most to the formation of SB$_{18}$. The small splitting actually merges the peaks due to the fact that each excited state ``prefers" ionization to a specific final ionic state. For instance, ionization of the $2p^5(^2P_{3/2})4d[3/2]$ (20.7~eV) state leads to $J_f=3/2$ in the final ion, while ionization of the $2p^5(^2P_{1/2})4d[3/2]$ (20.8~eV) state produces \hbox{$J_f=1/2$}.  Consequently, the resulting peaks in the spectrum center at the same energy. This fact somewhat justifies employing the non\-relativistic RMT model, which does not take into account the splitting. 

The pathways leading to excitation of the intermediate $2p^5 5s$ and $2p^5 6s$ resonances are found to play only a secondary role (see white bars in Fig.~\ref{Fig3}(a)). This is (i)~consistent with the propensity rule that the absorption of an XUV photon should preferably excite a state with increased orbital angular momentum~\cite{PRA-Fano-1985, PRL-Busto-2019} and (ii)~ionization cross sections from $d$-states to an $f$-wave \hbox{($d\rightarrow f$)} are by an order of magnitude larger than the $s\rightarrow p$ and \hbox{$d\rightarrow p$} cross sections near the threshold. 

The result of the presence of the two resonances is that the shape of the photo\-electron spectrum corresponding to the UT-RABBIT line (SB$_{18}$) strongly depends on the precise driving wavelength, as shown in Fig.~\ref{Fig3}(c). We can clearly observe a transition from a structure characterized by a single peak centered at high energies ($\lambda=1004$~nm; purple, dash-dotted line), to a double peak structure (\hbox{$\lambda=1006$}~nm; black, solid line), and finally to a larger peak at lower energies ($\lambda=1008$~nm; orange, dashed line). This evolution reproduces the experimental data presented in Fig.~\ref{Fig2}, panels \hbox{(b)-(e)}.

\begin{figure}
\centering
\centering 
\includegraphics[width=0.47\textwidth]{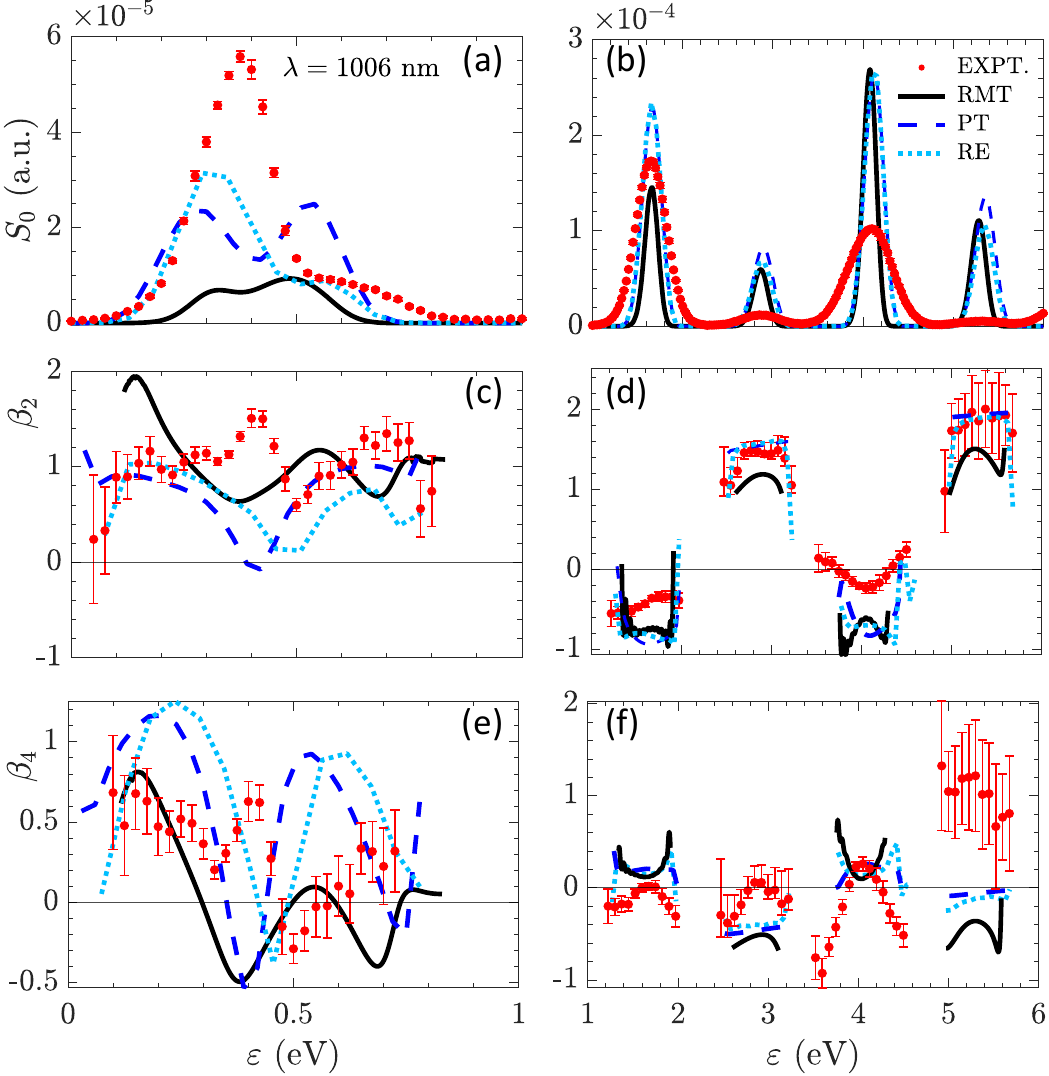} 
\caption{Parameters for the delay-averaged photo\-electron emission at an IR wavelength of 1006~nm:  (a,b): spectrum; (c,d): anisotropy parameter~$\beta_2$; (e,f): anisotropy parameter~$\beta_4$. The panels on the left column show the near-threshold region on an extended scale. The red points with error bars represent the experimental data, while the curves correspond to the theoretical predictions from RMT (solid black line), PT (blue dashed line), and RE (blue dotted line). In the gaps on the energy axis in panels~(d) and~(f), the signal is too low for a meaningful  extraction of the parameters. The experimental data for the spectra were scaled to provide a visual fit to the theoretical predictions.} 
\label{Fig4}
\end{figure}

For the same driving IR field wavelength of 1006~nm, the delay-averaged spectrum and the angular anisotropy parameters $\beta_2$ and $\beta_4$ were then extracted. They are presented in Fig.~\ref{Fig4} together with the values calculated by the three methods described above. The experimental data for the spectra were scaled to provide a good visual fit to the theoretical predictions. All theoretical curves were shifted to match the central energy position of the first main line (ML$_{19}$) in order to compensate for the Stark shift and make the comparison transparent. Note that the predicted Stark shift is different in the various models (zero in PT, 0.05~eV in~RE, and 0.08~eV in~RMT). The RMT prediction of the largest value for the Stark shift indicates stronger coupling to the discrete part of the spectrum compared to the other theories.

For electrons with energy above 1~eV (panel (b) of Fig.~\ref{Fig4}), all theoretical spectra are in a good agreement.  In the near-threshold region (panel~(a)), however, significant differences are seen due to the role of the discrete states involved in the process. All models exhibit a doublet structure. As discussed above, this doublet originates primarily from the $2p^5 4d$ and $2p^5 5d$ discrete states acting as intermediate steps in the two-color photo\-ionization process.  However, the detailed energy dependence of the doublet is different in each model. Specifically, the RE predictions are  closest to the experimental data shown in this panel, while PT exhibits a structure closer to the one seen in Fig.~\ref{Fig2}(c) for 1006~nm and RMT is closer to the one for 1001~nm (cf.\ Fig.~\ref{Fig2}(b)).

The principal reasons for the apparent discrepancy in the shape of the UT-RABBIT peak between the experimental data shown in Fig.~\ref{Fig2}(c) and Fig.~\ref{Fig4}(a) are the uncertainty in the determination of the central wavelength and the different distribution of the spectrum of the driving pulse, as well as a possible different IR intensity. 
 
The delay-integrated values of $\beta_2$ and $\beta_4$ defined according to Eq.~(\ref{eq:PADaveraged}) are shown together with the predictions of the various theoretical models in  panels \hbox{(c)-(f)} of Fig.~\ref{Fig4}. 
For the AT-RABBIT lines SB$_{20}$ and SB$_{22}$, there is good agreement in the retrieved $\beta_2$ from the experimental data and all theories (see panel (d)).
It is also remarkable that all theories show a significant difference between the calculated $\beta_2$ for the main lines ML$_{19}$ and ML$_{21}$ and the experimental values. Moreover, within the error bars, the latter are much closer to the single-photon values of $\beta_2=-0.5$ for ML$_{19}$ and $-0.16$ for ML$_{21}$~\cite{Codling1976}. This suggests that the IR intensity, and hence the contribution from third-order terms to the modulation in the main lines, may have been overestimated in the modelling.
In general, the agreement is not very good for~$\beta_4$ (panels (e) and~(f)). We note, however, that the determination of $\beta_4$ is less reliable, since its extraction from the experimental data is more challenging than that of~$\beta_2$. 

For the UT-RABBIT sideband SB$_{18}$ (panel (c)), the extracted parameters vary significantly through the line, because several discrete states affect the results to different extent. Note that a resonance often appears in the angular anisotropy parameters in a different way and possibly even at a slightly different energy than for the integrated spectrum~\cite{Grum_2005}. For the line-integrated anisotropy parameters, we found $\beta_2=1.2$ (expt.), \hbox{$\beta_2=0.9$} (RMT), $\beta_2=0.7$ (RE), and $\beta_2=0.6$ (PT).

All models predict a strong energy dependence in both~$\beta_2$ and~$\beta_4$ across the line in the low-energy region $0.2-0.8$~eV.  While significant structure is also seen in the experimental data, the details are clearly different. Recall that contributions from photo\-electron wave packets initially created via the $2p^5 4d$ and $2p^5 5d$ intermediate states overlap in this energy range. Together with the need to average the experimental data over windows of energies and space to extract the $\beta_n$ parameters, this makes reliable calculations extremely challenging and a comparison with experiment difficult. 
\newpage
\subsection{Delay-resolved data}
\begin{figure}[h]
\centering 
\includegraphics[width=0.42\textwidth]{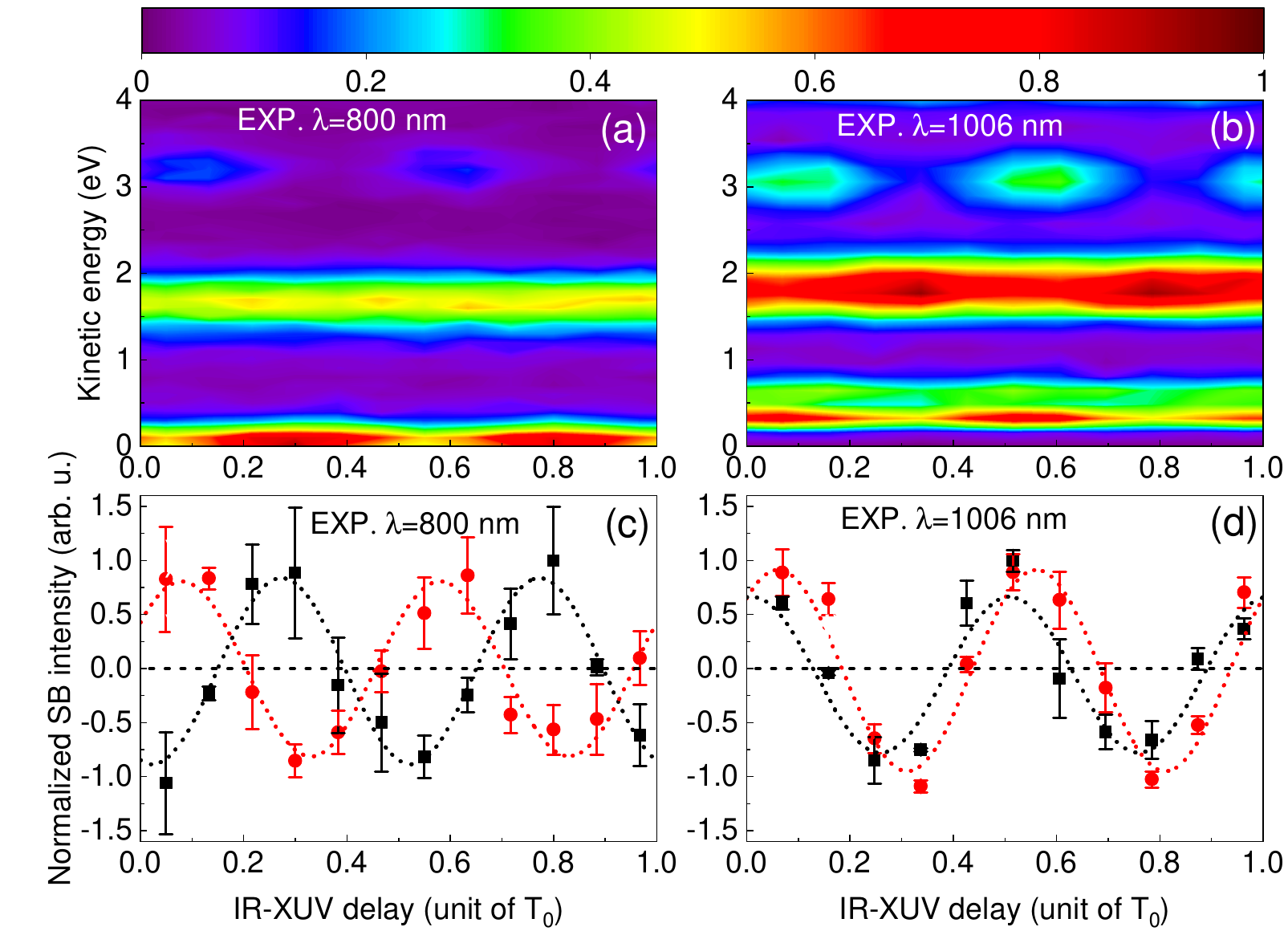}
\includegraphics[width=0.42\textwidth]{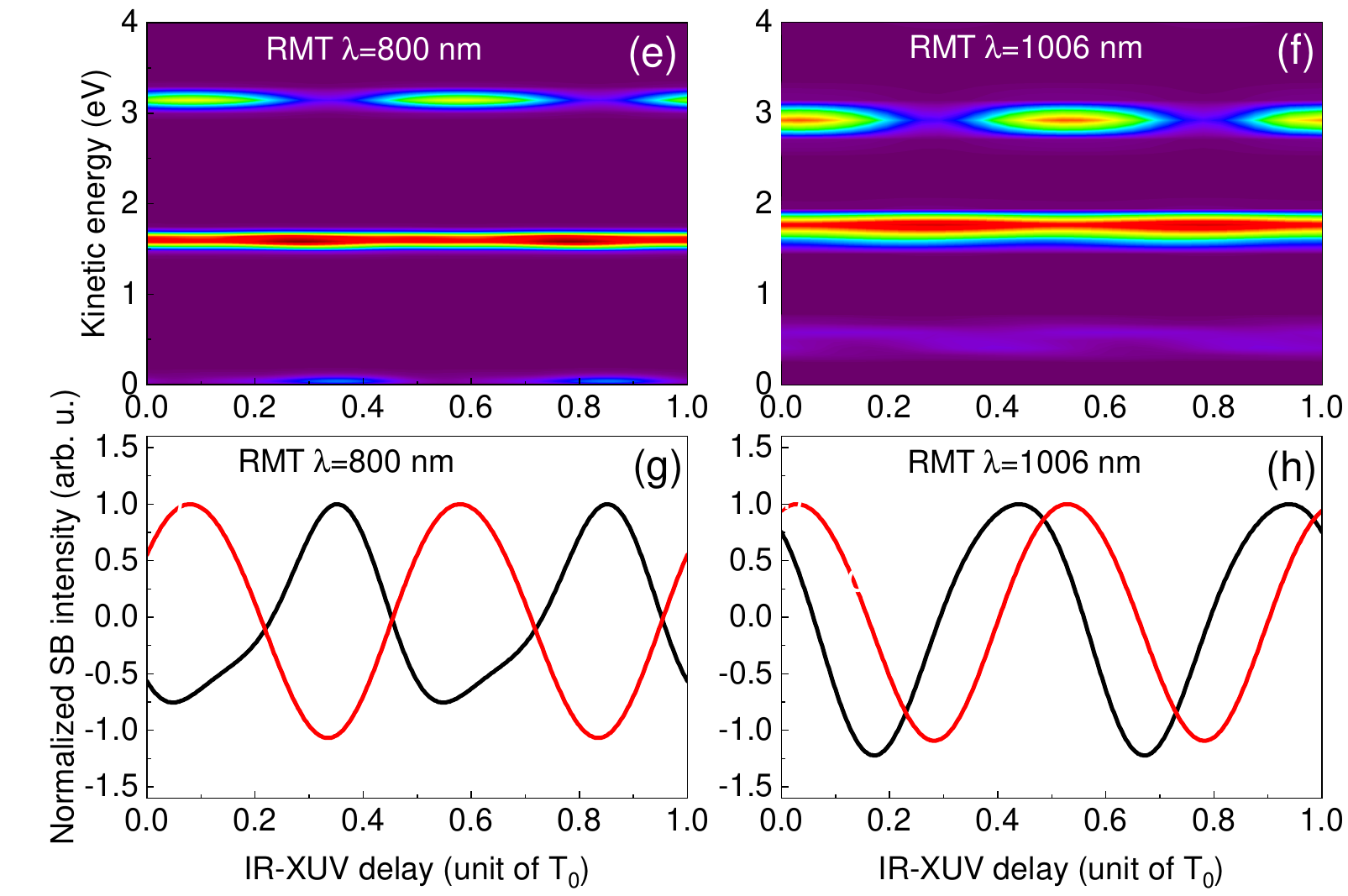}
\includegraphics[width=0.42\textwidth]{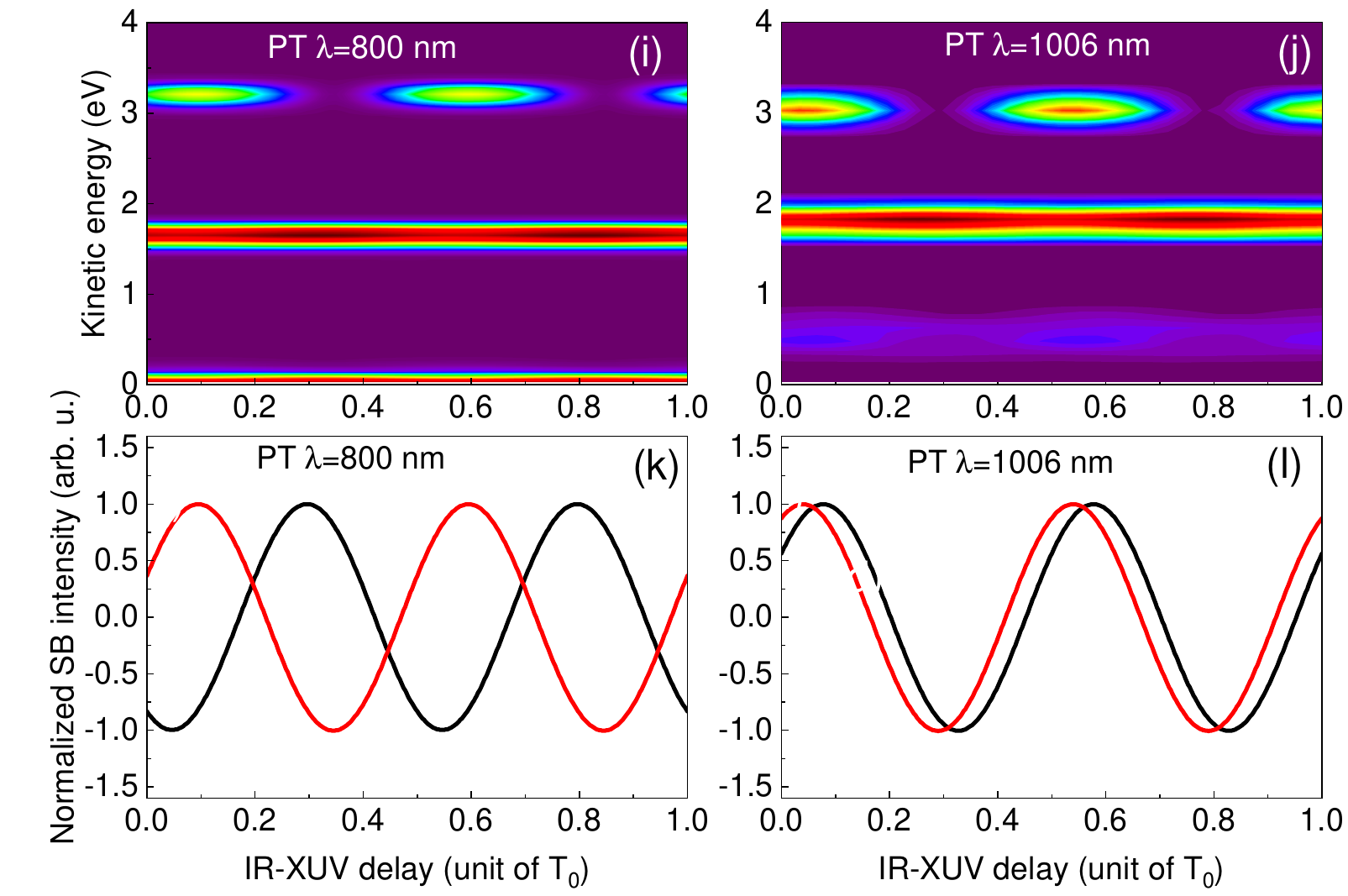}
\includegraphics[width=0.42\textwidth]{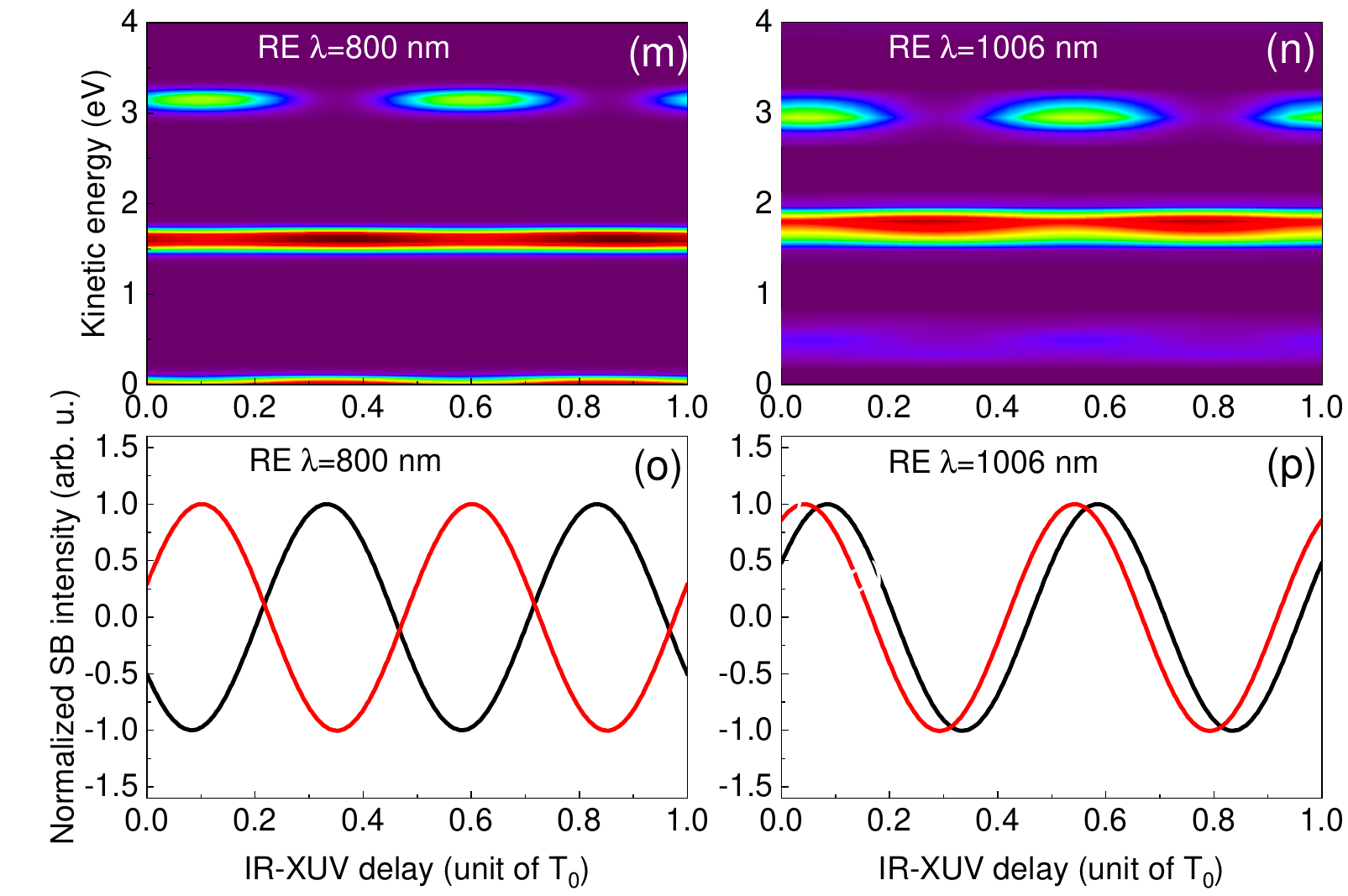}
\caption{\textcolor{black}{\hbox{RABBIT} traces (a,b,e,f,i,j,m,n) in the range $0-4$~eV and oscillations of the UT-RABBIT (black line) and first AT-RABBIT (red lines) sidebands (panels c,d,g,h,k,l,o,p) extracted from the experimental data (a-d), RMT (e-h), PT (i-l), and RE (m-p) models for $\lambda=800$~nm (left panels) and $\lambda=$1006~nm (right panels). The oscillations of the sideband signals are plotted relative to the average intensity of the sideband, which we set as the zero line for the normalized intensity. The dotted lines in panels (a) and (b) are sinusoidal fits to the experimental data.}}
\label{Fig5}
\end{figure}

We now focus the discussion on the comparison of the experimental and calculated data for the first sideband above the ionization threshold (UT-RABBIT). 
\hbox{RABBIT} traces as a function of the delay $\tau$ between the XUV and IR fields in the low kinetic energy region are shown for $\lambda=800$~nm and $\lambda=1006$~nm in Fig.~\ref{Fig5}(a) and~(b), respectively. The intensity oscillations of the first two sidebands can be fitted by cosine functions (see Fig.~\ref{Fig5}(c,d)), from which the RABBIT phases can be extracted according to Eq.~(\ref{eq:Klaus}). The phase difference between the UT-RABBIT sideband and the first AT-RABBIT sideband is remarkably larger for the 800~nm wavelength driver ($\Delta\phi=2.39\pm0.16$~rad) compared to the 1006~nm case ($\Delta\phi=-0.65\pm0.27$~rad).
\textcolor{black}{The large phase shift for the first wavelength is well reproduced by all the simulation models, as shown in Fig.~\ref{Fig5}, although with some difference in the exact value (see Table~\ref{Table2}). The phase shift for the $1006$-nm-case is partially reproduced only by the RMT simulations, while the PT and RE model predict a small difference but with the opposite sign to the experimental one. We attribute the differences observed for the three different models in the second case to the description of the large number of states below the ionization threshold.}


\begin{table}[h]
  \centering
\begin{tabular}{| c | c | c | c |  p{8cm} |}
    \hline \multicolumn{4} {|c|} {\bfseries $\lambda=800$ nm}\\ \hline
     & $\phi(\mathrm{UT})$ (rad) & $\phi(\mathrm{AT})$ (rad) &  $\Delta\phi$ (rad)  \\ \hline
     Exp. & $3.46\pm0.08$ & $1.07\pm0.08$ & $2.39\pm0.16$   \\ \hline
     RMT &  $4.25$ & $1.04$ & $3.21$   \\ \hline
     PT &  $3.72$ & $1.19$ & $2.53$  \\ \hline
     RE & $4.21$ & $1.32$  & $2.89$ \\\hline\hline
       \hline \multicolumn{4} {|c|} {\bfseries $\lambda=1006$ nm}\\ \hline
     & $\phi(\mathrm{UT})$ (rad) & $\phi(\mathrm{AT})$ (rad) &  $\Delta\phi$ (rad)  \\ \hline
     Exp. & $0.12\pm0.15$ & $0.77\pm0.12$ & $-0.65\pm0.27$   \\ \hline
     RMT &  $-0.92$ & $0.39$ & $-1.31$   \\ \hline
     PT &  $0.97$ & $0.50$ & $0.47$  \\ \hline
     RE & $1.07$ & $0.53$  & $0.54$ \\\hline\hline
\end{tabular}
  \caption{\textcolor{black}{Phase $\phi$ extracted using sinusoidal fits from the experimental data and simulations using the three different models for the UT and AT sidebands and for the two driving wavelengths $\lambda=800$~nm and $\lambda=1006$~nm. The phase difference $\Delta\phi=\phi(\mathrm{UT})-\phi(\mathrm{AT})$ is also reported. The error bars for the experimental data derived directly from the fitting algorithm.}}\label{Table2}
\end{table}

\begin{figure}[h!]
\centering
\includegraphics[width=0.42\textwidth]{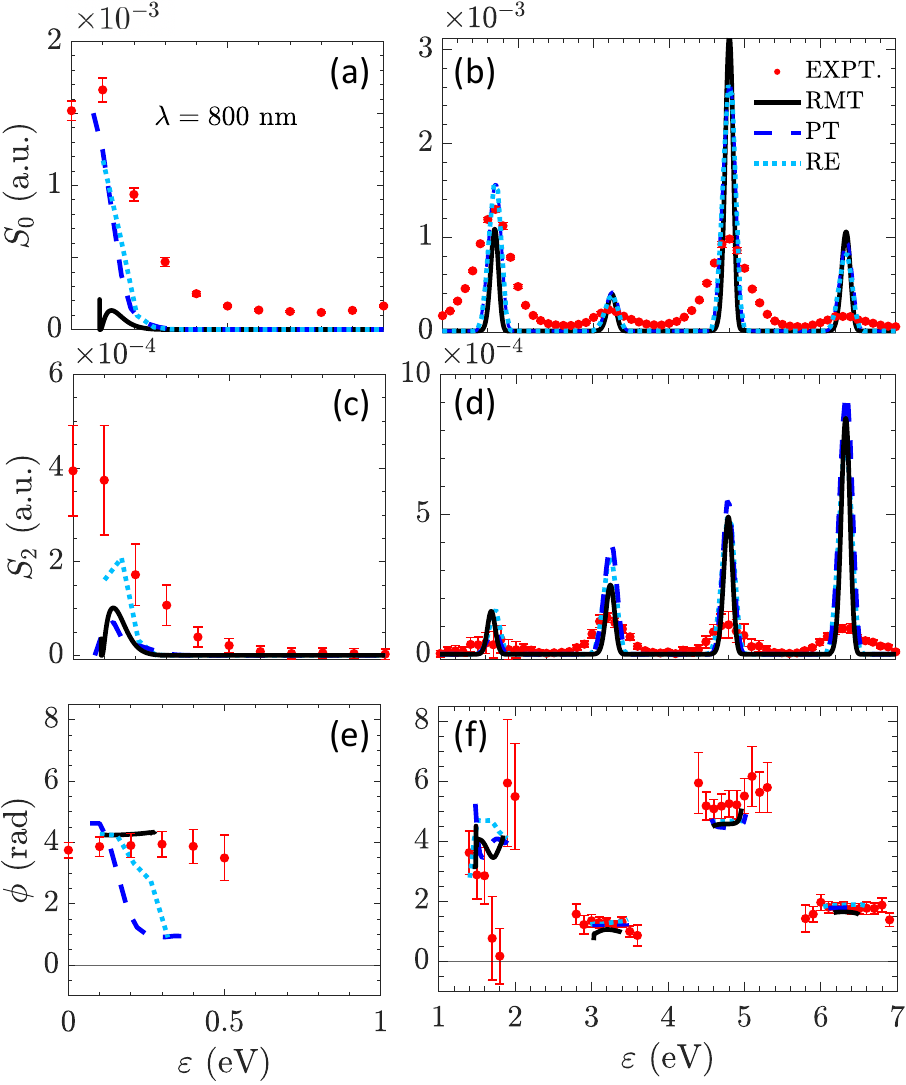}
\caption{Delay-averaged parameters for photo\-electron emission at an IR wavelength of 800~nm.  (a,b): spectrum; (c,d): oscillation amplitude; (e,f): the corresponding phase extracted according to Eq.~(\ref{eq:Klaus}). The panels on the left column show the near-threshold region on an extended scale. Since the phase extraction only makes sense near the peaks of the emission spectra, there are some gaps in the energy between the peaks.
} 
\label{Fig6}
\end{figure}

\begin{figure}[h!]
\centering 
\includegraphics[width=0.47\textwidth]{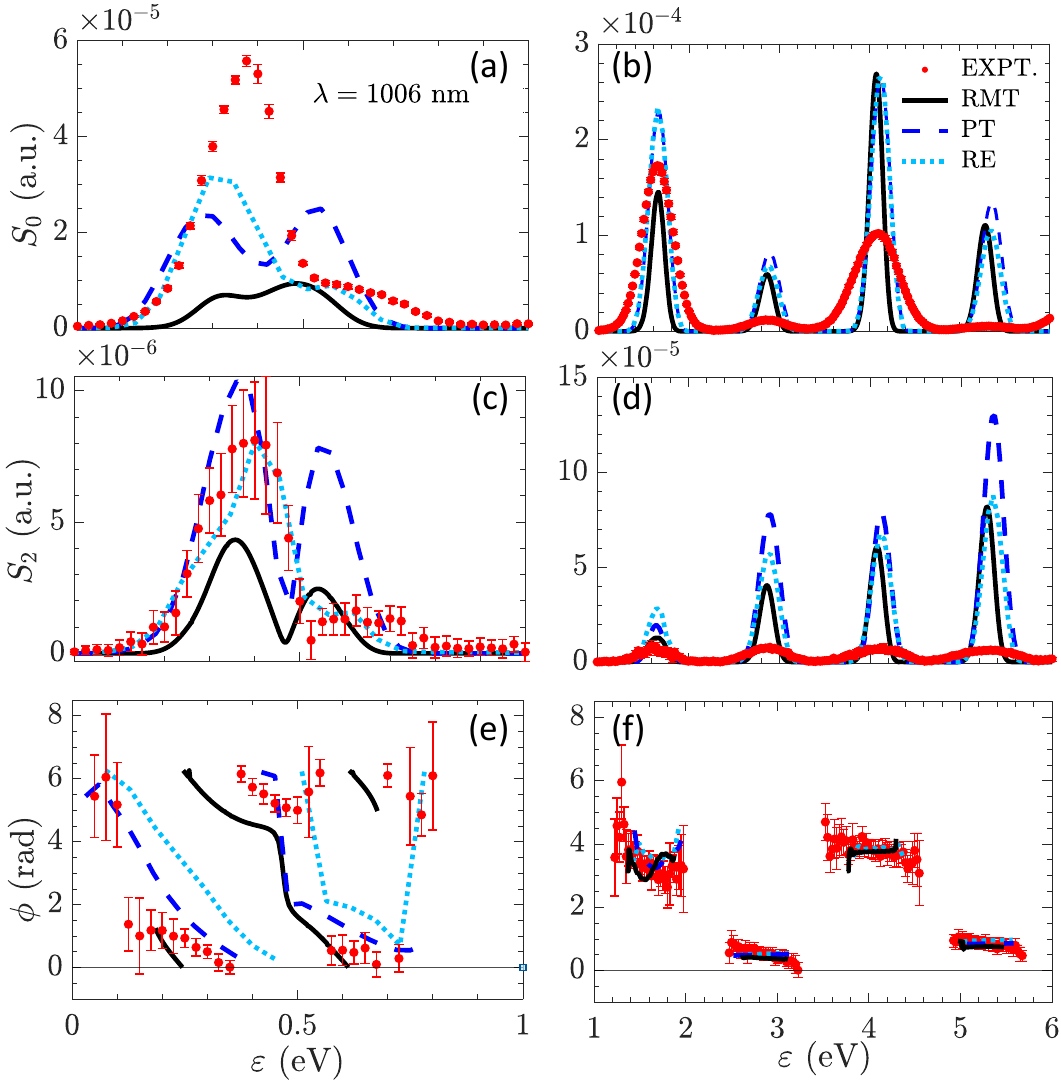}
\caption{Delay-averaged parameters for photo\-electron emission at an IR wavelength of 1006~nm.  (a,b): spectrum; (c,d): oscillation amplitude; (e,f): the corresponding phase extracted according to Eq.~(\ref{eq:Klaus}). The panels on the left column show the near-threshold region on an extended scale. Since the phase extraction only makes sense near the peaks of the emission spectra, there are some gaps in the energy between the peaks.}
\label{Fig7}
\end{figure}

\clearpage
Figures~\ref{Fig6} and ~\ref{Fig7} present comparison of $S_0$, $S_2$ and $\phi$ extracted from the experimental data and the simulations.
Starting with Fig.~\ref{Fig6}, we notice good qualitative agreement between all three theories in $S_0$ for the energy range $1-7$~eV (panel~(b)), while there are differences near threshold (panel~(a)).  Note that both the theoretical and the experimental data are cross-normalized between the panels with a single normalization factor.  The latter was chosen to give a reasonable visual fit to the experimental data. 
With the present choice, the integrals under the peaks near 3.4~eV and 4.9~eV are about the same in experiment and the three theories, while the theories underestimate the peak near 1.8~eV and overestimate the one near 6.4~eV.  At very low energies, PT and RE reproduce at least qualitatively the fall-off with increasing photoelectron energy, which is due to the peak reaching below ionization threshold.  On the other hand, RMT predicts this peak to be further below the threshold than the other two theories. This results in a very small value of~$S_0$ in the $0-1$~eV range.
Regarding~$S_2$, there is better overall agreement among the three sets of theoretical predictions than for~$S_0$, with the largest differences occurring in the near-threshold region (panel~(c)). In the $1-7$~eV range (panel~(d)), the theories agree well with each other and, after accounting for the different widths of the peaks, the agreement with the experimental predictions also improved compared to~$S_0$.
Finally, the experimental \hbox{RABBIT} phase is well reproduced by all theories in the energy range 3-7~eV (panel (f)), for the peaks around 3.4~eV, 4.9~eV, and 6.4~eV. 

The complex phase evolution for the near-threshold UT-RABBIT sideband and the first mainline around 1.8~eV (see Fig.~\ref{Fig6}(e,f)) indicates the importance of the under-threshold bound states that complicate the situation.  Interestingly, while clearly underestimating the height of the UT-RABBIT sideband, RMT gives a flat phase, whereas PT and RE show a rapid decrease in this energy region (panel~(e)). All theories predict a nearly flat phase across the 1.8~eV peak, while the experimental data, albeit with large error bars, suggest a rather strong energy dependence.  

Figure~\ref{Fig7} shows the corresponding results for a central wavelength of $\approx 1006$~nm.  Panels~(a) and~(b) are the same as in Fig.~\ref{Fig4}.  

In the near-threshold region, all theories predict a doublet structure in $S_2$ (panel~(c)), similar to that seen in $S_0$ in Fig.~\ref{Fig4}(a) and Fig.~\ref{Fig7}(a). Here the RE model is closest to the experimental data regarding the energy dependence. 
\textcolor{black}{We attribute this agreement to the description of the fine-structure splitting. Even though RMT is an ab-initio method, it describes the process in the LS-scheme. In our case, however, the 3d states are highly mixed due to the fine-structure interaction. This is best described in the \emph{jK}-coupling scheme adopted in the RE approach.}
It should be pointed out that the data presented in Figs.~\ref{Fig4} and~\ref{Fig7} were taken under slightly different experimental conditions than those presented in Fig.~\ref{Fig2}. For this reason, care should be taken in evaluating the quality of the comparison. Note, in particular, that the height of the experimental sidebands in panel~(b) is relatively low, suggesting that the IR intensity in the calculations was significantly higher than in the experiment. This issue is also clearly seen in panel~(d), where the agreement between the theories is at a level of about a factor of~2, while the calculated oscillation amplitudes are much larger than in the experimental data.

Looking at the \hbox{RABBIT} phase, we see excellent agreement between all theoretical predictions and also experiment in the $1-7$~eV region (panel~(f)), with a jump of $\approx \pi$ between consecutive peaks and a nearly constant phase across the peaks. In this case, however, the \hbox{RABBIT} phase varies strongly across the near-threshold structure (panel~(e)).  The rapid change seen in the experimental data as a function of the ejected-electron energy is to a large extent even quantitatively reproduced by all theories, with RMT being the closest to the experimental data.
The principal reason for this rapid variation is, once again, the presence of at least two discrete bound states under the ionization threshold that affect different energy regions of the UT-RABBIT sidebands. The interplay between the contributions of the below-threshold resonances leads to large variations in the phase of the two-color photoionization peak, which are qualitatively reproduced by the theoretical models.

\subsection{Triple-differential data}

\begin{figure}[t!]
\centering 
\includegraphics[width=0.47\textwidth]{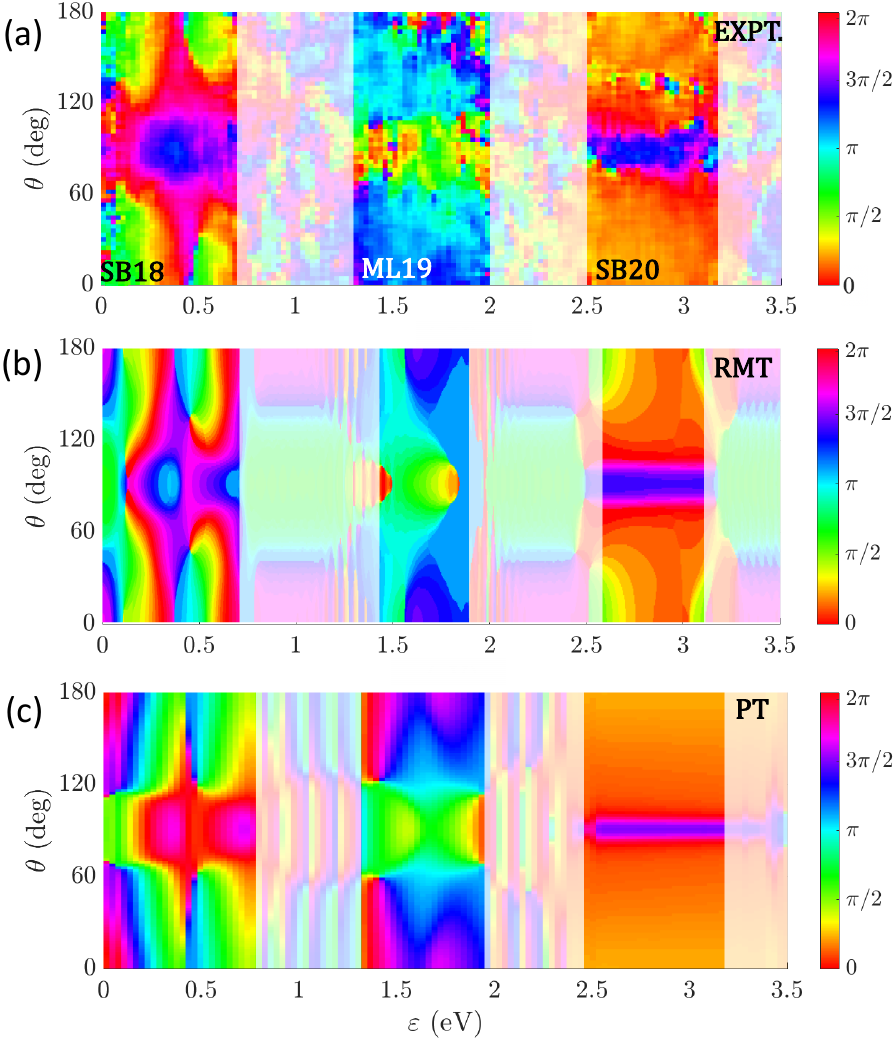}
\caption{Angle-resolved atomic phases for the driving field with $\lambda=1006$~nm: (a)~experiment; (b)~RMT; (c)~PT. 
In the energy windows with the faint colors, the 
signal is too small to allow a meaningful comparison.}
\label{Fig8}
\end{figure}

In this section we present the energy- and angle-resolved \hbox{RABBIT} phase that we can extract from the experimental and simulated side\-band oscillations according to Eq.~(\ref{eq:PAD}). 
As indicated in the discussion of the angle-integrated phases in Figs.~\ref{Fig6} and~\ref{Fig7}, the phase is rather insensitive to details in the peak heights, i.e., the particular IR intensity, as long as one remains in the (second-order) perturbative regime.
The evolution of the \hbox{RABBIT} phase $\phi(\theta)$ over the entire range of angles for the energy interval \hbox{$0-3.5$~eV} is presented in Fig.~\ref{Fig8} for the driving field with $\lambda=1006$~nm.  Overall, there is very satisfactory agreement between the experimental map and that predicted by the RMT model, especially in light of the fact that SB$_{18}$ in the RMT calculation is narrower than in the experiment. 
The key characteristics of the experimental data are also reproduced by the PT approach. The same is true for the RE map (not shown), which is very similar to that obtained with PT. 

The experimental map indicates rapid phase variations around $\theta=90^{\circ}$ as a function of energy.  Note that the cross section here is minimal, since $\beta_2>0$. This dependence is reproduced well in the calculations. Both the experimental data and the theoretical predictions again exhibit a rich structure in the $0.4-0.5$~eV energy region, where the pathways through the $2p^5 4d$ and $2p^5 5d$ resonances provide comparable contributions. The evolution in the proximity of these resonances is well reproduced in the calculations. 

\bigskip

\section{Conclusions} \label{Conclusions}
We have presented a joint experimental and theoretical investigation of atto\-second photo\-electron inter\-ferometry around the ionization threshold of neon. Experimental data were acquired using atto\-second pulse trains driven by two different driving fields with wavelengths of \hbox{$\lambda \approx 800$~nm} and \hbox{$\lambda\approx 1006$~nm}. For the latter configu\-ration, fine-tuning of the driving laser wavelength in the range $\rm 1001~nm\leq\lambda\leq1019~nm$ made it possible to change the relative weight of closely-spaced inter\-mediate resonances.
The angular-resolved data indicate how the presence of resonances is affecting the photo\-electron angular distributions of the sidebands close to the ionization threshold.
\textcolor{black}{The presence of resonances below the ionization threshold can affect the photoionization phase in RABBIT measurements depending on the specific resonance and photon energy. In the case of atoms, the contributions of the different resonances can be disentangled by exploiting the tunability of the laser and the resolution of the photoelectron spectrometer. In molecular systems, isolating the contributions of the different resonances is a major challenge due to the complex electronic and vibrational structure. Their influence on the RABBIT spectra will require the development of accurate theoretical methods \cite{SciAdvBorras2023,BORRAS2024109033} to describe the dense energy distribution around the ionization threshold.}
\vspace{-3.0truemm}

\section*{Acknowledgments} 
The experimental part of this work was supported by the European Union’s Horizon 2020 project 641789 MEDEA.
D.B.\ acknowledges support from the Swedish Research Council grant 2020-06384. G.S.\ acknowledges support by FRIAS. I.M.\ and G.S.\ acknowledge support by the BMBF project 05K19VF1 and the Georg H.\ Endress Foundation.
The theoretical part of this work was supported by the NSF under OAC-1834740, PHY-2110023, PHY-2408484 (K.B.), OAC-2311928 (K.R.H.\ and K.B.), the ACCESS allocation PHY-090031, and the Frontera Pathways allocation PHY20028.
The RMT calculations were performed on Stampede-2 and Frontera at the Texas Advanced Computing Center.

\bibliography{wwhhg}

\end{document}